\documentclass[aps,showpacs,prl,twocolumn,superscriptaddress]{revtex4-1}
\usepackage[pdftex]{graphicx}
\usepackage{slashbox}
\usepackage{ulem}

\newcommand{\ri}{\it i}
\begin{document}

\date{\today}
\title{Testing the validity of the strong spin-orbit-coupling limit for octahedrally coordinated iridates in a model system Sr$_3$CuIrO$_6$}

\author{X. Liu}
\email[electronic address: ]{xliu@bnl.gov and xliu@iphy.ac.cn}
\affiliation{Condensed Matter Physics and Materials Science Department, Brookhaven National Laboratory, Upton, New York 11973, USA\\}
\author{Vamshi M. Katukuri}
\affiliation{Institute for Theoretical Solid State Physics, IFW Dresden, Helmholtzstrasse 20, 01069 Dresden, Germany\\}
\author{L. Hozoi}
\affiliation{Institute for Theoretical Solid State Physics, IFW Dresden, Helmholtzstrasse 20, 01069 Dresden, Germany\\}
\author{Wei-Guo Yin}
\affiliation{Condensed Matter Physics and Materials Science Department, Brookhaven National Laboratory, Upton, New York 11973, USA\\}
\author{M. P. M. Dean}
\affiliation{Condensed Matter Physics and Materials Science Department, Brookhaven National Laboratory, Upton, New York 11973, USA\\}
\author{M. H. Upton}
\affiliation{Advanced Photon Source, Argonne National Laboratory, Argonne, Illinois 60439, USA\\}
\author{Jungho Kim}
\affiliation{Advanced Photon Source, Argonne National Laboratory, Argonne, Illinois 60439, USA\\}
\author{D. Casa}
\affiliation{Advanced Photon Source, Argonne National Laboratory, Argonne, Illinois 60439, USA\\}
\author{A. Said}
\affiliation{Advanced Photon Source, Argonne National Laboratory, Argonne, Illinois 60439, USA\\}
\author{T. Gog}
\affiliation{Advanced Photon Source, Argonne National Laboratory, Argonne, Illinois 60439, USA\\}
\author{T. F. Qi}
\affiliation{Center for Advanced Materials, University of Kentucky, Lexington, Kentucky 40506, USA\\}
\affiliation{Department of Physics and Astronomy, University of Kentucky, Lexington, Kentucky 40506, USA\\}
\author{G. Cao}
\affiliation{Center for Advanced Materials, University of Kentucky, Lexington, Kentucky 40506, USA\\}
\affiliation{Department of Physics and Astronomy, University of Kentucky, Lexington, Kentucky 40506, USA\\}
\author{A.  M. Tsvelik}
\affiliation{Condensed Matter Physics and Materials Science Department, Brookhaven National Laboratory, Upton, New York 11973, USA\\}
\author{Jeroen van den Brink }
\affiliation{Institute for Theoretical Solid State Physics, IFW Dresden, Helmholtzstrasse 20, 01069 Dresden, Germany\\}
\author{J. P. Hill}
\affiliation{Condensed Matter Physics and Materials Science Department, Brookhaven National Laboratory, Upton, New York 11973, USA\\}

\begin{abstract}
The electronic structure of Sr$_3$CuIrO$_6$, a model system for the 5{\it d} Ir ion in an octahedral environment, is studied through a combination of resonant inelastic x-ray scattering (RIXS) and theoretical calculations. RIXS spectra at the Ir L$_3$-edge reveal an Ir $t_{2g}$ manifold that is split into three levels, in contrast to the expectations of the strong spin-orbit-coupling limit. Effective Hamiltonian and $ab~inito$ quantum chemistry calculations find a strikingly large non-cubic crystal field splitting comparable to the spin-orbit coupling, which results in a strong mixing of the $j_{\mathsf{eff}}=\frac{1}{2}$ and $j_{\mathsf{eff}}=\frac{3}{2}$ states and modifies the isotropic wavefunctions on which many theoretical models are based.
\end{abstract}


\maketitle

A new type of Mott physics has recently been discussed in the $5d$ transition metal oxides\cite{Mott1,Mott2,Mott3}. In contrast with the more familiar $3d$ Mott insulators, for which new physical phenomena originate from the large on-site Coulomb interaction, {\it U}, in $5d$ transition metal oxides the strong electron correlation has been argued to be driven by a large spin-orbit coupling (SOC). Iridates are an important example in this category and have attracted much attention recently\cite{Mott1,Mott2,Mott3,honeycomb1,honeycomb2,honeycomb3,pyrochlore2,pyrochlore3,hyperkagome2Y,hyperkagome2L,hyperkagome2M,hyperkagome2Yi,SC1,SC214,Gang1,Gang2,sizeU1,sizeU2,mag327}. In particular, iridates in which the iridium atom sits in an octahedral oxygen cage have been classified as narrow-band Mott insulators. It is argued that strong SOC splits the occupied $t_{2g}$ orbitals into two sets of narrow bands corresponding to $j_{\mathsf{eff}}=\frac{1}{2}$ and $j_{\mathsf{eff}}=\frac{3}{2}$ states. For the important case of the Ir$^{4+}$, $5d^5$ ion, this puts one electron in the $j_{\mathsf{eff}}=\frac{1}{2}$ state and even a small $U$, then gives rise to Mott insulating behavior. Further, because of the isotropic nature of the resulting wavefunctions\cite{Mott3}, this has important consequences for the magnetic superexchange interactions and has led to predictions of an array of interesting physics, including quantum spin liquids\cite{honeycomb2,pyrochlore2,hyperkagome2Y,hyperkagome2L,hyperkagome2M,hyperkagome2Yi}, topological insulators\cite{Mott2,honeycomb1} and superconductivity\cite{SC214}.

The strong SOC limit assumes local cubic symmetry with perfect IrO$_6$ octahedra. In reality, all of the proposed physical realizations of these models have non-ideal octahedra. Examples include the pyrochlores $R_2$Ir$_2$O$_7$ ({\it R} = Y, Sm, Eu and Lu), the hexagonal lattice systems $A_2$IrO$_3$ ({\it A} = Li, Na), the hyperkagome lattice compound Na$_4$Ir$_3$O$_8$ and the square lattice compounds (Ba,Sr)$_2$IrO$_4$. To date these have mostly been treated in the strong SOC limit. There are signs though, that this may not always be appropriate. For example Na$_2$IrO$_3$ has been thought to be a realization of the Kitaev model on a honeycomb lattice\cite{honeycomb2}. However,  It was found recently to be antiferromagnetically ordered in a zigzag pattern\cite{M213,N213}, contradicting theoretical predictions based on the strong SOC limit. This is suggestive of significant non-cubic crystal fields and indeed very recently theorists have begun to consider this question \cite{TCF1, TCF2}. Thus, understanding the interplay of SOC and non-cubic crystal fields and experimentally determining their size are crucial requirements for any realistic model for the iridates.
\begin{figure}[ht]
\includegraphics[width=0.48\textwidth]{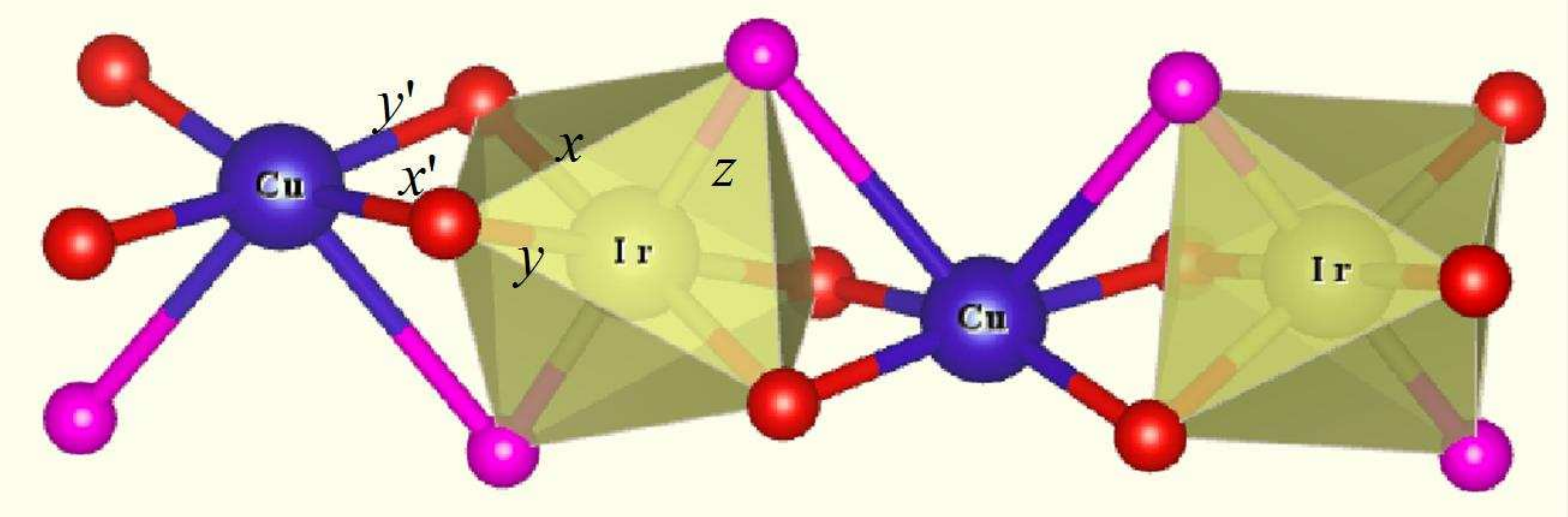}
\caption{(Color online) The Cu-IrO$_6$-Cu-IrO$_6$ chain of Sr$_3$CuIrO$_6$. The local coordinates for the IrO$_6$ octahedra and the CuO$_6$ prism are labeled with ($x,y,z$) and ($x^\prime,y^\prime$) respectively.}
\label{fig1}
\end{figure}

In this work, we experimentally determine the spin-orbit coupling and crystal field splitting in a model iridate, Sr$_3$CuIrO$_6$\cite{str3116}, which has only small distortions of the IrO$_6$ octahedra. We find that there is strong mixing of the $j_{\mathsf{eff}}=\frac{1}{2}$ and $j_{\mathsf{eff}}=\frac{3}{2}$ states and that, even in this case, it is not appropriate to work in the strong SOC limit. An important consequence of this mixing is that the wavefunctions are no longer isotropic and the magnetic superexchange interactions are modified away from the conventional picture. These results suggest that the conventional $j_{\mathsf{eff}}=\frac{1}{2}$ Mott insulator picture of many of the currently interesting octahedrally coordinated iridates needs to be revisited in light of these effects.

Before presenting the results in detail, we first note that there are several unique features of this study of Sr$_3$CuIrO$_6$ that make it ideal for probing these effects. First, the oxygen octahedra are well separated from each other and therefore one does not need to consider hybridization effects between them\cite{hybri}. Second, the octahedra themselves only have relatively small distortions - the Ir-O bond lengths vary by less than 2\% (full details of the structure are provided below), so that non-cubic crystal fields are expected to be small. Third, we use the technique of resonant inelastic x-ray scattering to probe the energy and momentum dependence of the excitation spectrum. This technique has the advantage of probing only the orbital and magnetic excitations associated with the Ir sites, making interpretation of the spectra straightforward. This, combined with the fact that the iridium-derived electronic states are well localized, allows us to compare to the theoretical models with substantial precision. We observe six well-defined electronic transitions. By comparing with an effective model Hamiltonian and with $ab~initio$ theoretical calculations, we are able to derive a complete description of the electronic degrees of freedom. 

Sr$_3$CuIrO$_6$ is a quasi-one-dimensional material with a monoclinic structure\cite{str3116,mag3116}. Individual IrO$_6$ octahedra are linked by spin-1/2 Cu$^{2+}$ ions along one direction, forming a chain structure (fig. 1). Intersite hopping is largely suppressed due to this special structure: the $d_{x^2-y^2}$ orbital at the Cu site is essentially orthogonal to the $t_{2g}$ orbitals on the Ir site, resulting in a negligible intersite hopping integral. Even though the extended nature of the $5d$ orbitals tends to reduce correlation effects, this special arrangement puts Sr$_3$CuIrO$_6$ in the strongly localized regime.
\begin{figure}[ht]
\includegraphics[width=0.48\textwidth]{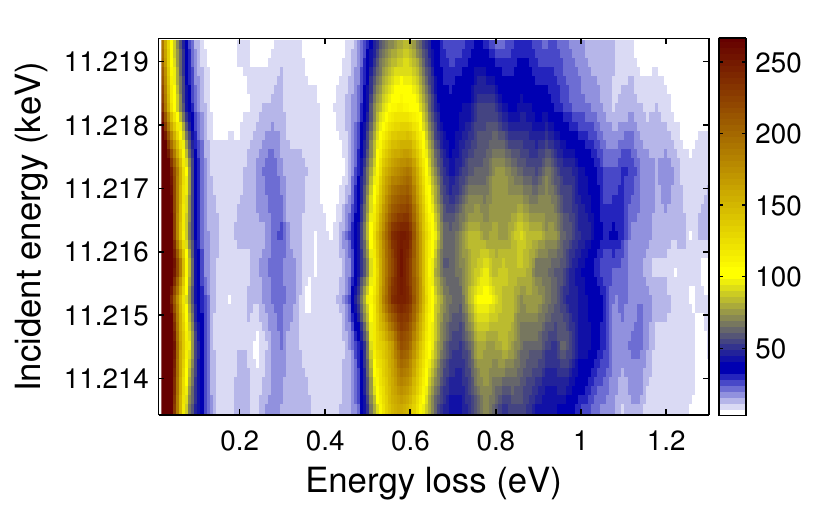}
\caption{(Color online) Incident x-ray energy dependence of the RIXS spectra from Sr$_3$CuIrO$_6$ at the Ir L$_3$ edge. Three low energy features at 0.28, 0.58 and 0.81 eV are observed. These show similar resonant behavior, each peaking at 11.216 keV.}
\label{fig2}
\end{figure}

Single crystal Sr$_3$CuIrO$_6$ was grown with self-flux techniques. In reciprocal space, the chain direction is parallel to Q$\approx$(1.98 0 1). In our measurements, Q points at $\zeta$(1.98 0 1)+(0 13 0) were selected along the chain direction. A single chain is shown in Fig.\,1 with the local coordinates for a single IrO$_6$ octahedra marked. The O-Ir-O bonds are straight and of similar length with a variation of less than $2\%$. The distortion of the IrO$_6$ octahedra is mainly rotations of the O-Ir-O bonds. The in-plane bonds along the $x$ and $y$ directions rotate towards each other, reducing the angle from 90 to 80 degrees, and the apical oxygens rotate towards the Cu atoms by $\sim 4$ degrees. We shall see that even these small distortions are sufficient to break the degeneracy of the t$_{2g}$ manifold and move the system out of the strong SOC limit.

The energy and momentum dependence of the excitation spectrum was studied using resonant inelastic x-ray scattering \cite{RIXS}. The experiments were carried out at the Ir L$_3$ edge for which dipole transitions excite and de-excite a $2p_{\frac{3}{2}}$ core-electron to the $5d$ orbitals. Due to the strong core-hole potential and localized nature of the Ir $5d$ electrons in Sr$_3$CuIrO$_6$, the RIXS process is dominated by local dipole transitions which lead to intra-site $d$-$d$ excitations, an important consideration for this work. The measurements were carried out at beamline 9-ID, Advanced Photon Source, in a horizontal scattering geometry. A Si(844) secondary monochromator and a R=2m Si(844) diced analyzer were utilized. The overall energy resolution of this setup was $\sim 45$meV (FWHM). All data were collected at 7 K.

The incident X-ray energy dependence of the RIXS spectra near the Ir L$_3$ edge is shown in Fig.\,2. Three features are observed at 0.28, 0.58 and 0.81 eV in the low energy region. Importantly, these three features show the same resonant behavior as function of incident x-ray energy, all resonating around 11.216 keV. This indicates that they all originate from initial $2p\rightarrow 5d$ transitions into the same unoccupied states within the Ir t$_{2g}$ manifold. The different energies arise from the fact that the $5d\rightarrow 2p$ de-excitation transitions create holes in different orbitals, leaving the system in different excited states at the end of the respective RIXS processes.
\begin{figure}[ht]
\includegraphics[width=0.48\textwidth]{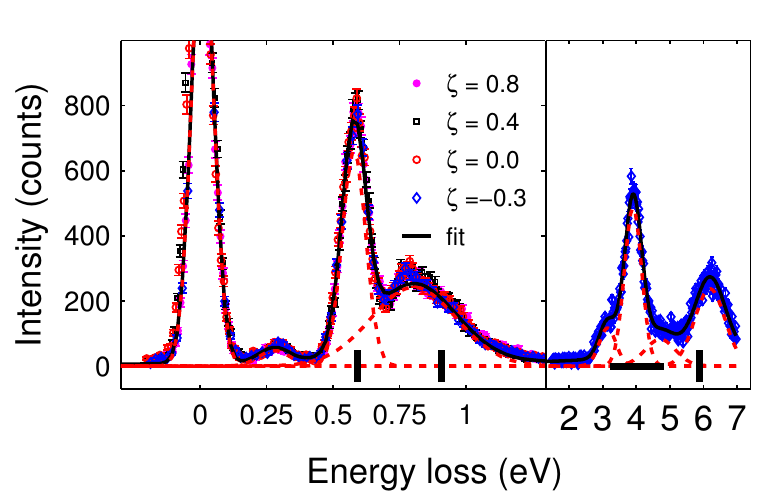}
\caption{(Color online) RIXS spectra at different Q = $\zeta$(1.98 0 1)+(0 13 0) points along the chain direction. The data at $\zeta$ = -0.3 are fit with multiple Gaussian peaks. The black solid line shows the total fitting result. Individual Gaussian peaks are shown with red dashed lines, which are centered at 0.28, 0.58, 0.81, 3.14, 3.90, 4.72 and 6.20 eV, respectively. The vertical and horizontal bars indicate the excitations listed in Table\,I from the MRCI+SOC calculations.}
\label{fig3}
\end{figure}

More details of these three excitations can be seen in Fig.\,3. The peak at 0.28eV, though sharing the same resonant behavior, has much less spectral weight. This excitation is most likely from impurity states associated with defects. At higher energy loss above 2eV, $d$-$d$ excitations into Ir $e_g$ orbitals and charge transfer excitations from oxygen 2$p$ orbitals are also observed. Here we focus on the 0.58 and 0.81 eV excitations, which are $d$-$d$ excitation within the iridium $t_{2g}$ manifold. These two peaks, especially the lower energy one, are fairly sharp. Importantly, they show no significant dispersion with momentum transfer (Fig.\,3). These observations are consistent with the narrow band picture and the very localized nature of the 5$d$ electrons in Sr$_3$CuIrO$_6$.

In the ideal case for which the  Ir atom sits within a regular oxygen octahedral cage, the $t_{2g}$ manifold is split into $j_{\mathsf{eff}} = \frac{1}{2}$ and $j_{\mathsf{eff}} = \frac{3}{2}$ states by the SOC\cite{Kim1}. In this effective 2 level system, only one $d$-$d$ transition is expected. Our observations clearly deviate from this ideal picture, and even without any modeling, already point to the importance of non-cubic crystal fields.

In order to understand these results, we first turn to an effective Hamiltonian to describe the local Ir $5d$ electrons. This will provide us with an intuitive description of the electronic structure. We will later see that this is consistent with a more sophisticated $ab~initio$ quantum chemistry calculation for this structure. We start with a local atomic multiplet Hamiltonian in hole representation,
\begin{equation}
H_{local} = \lambda\sum_{n}\vec{L}_{n}\cdot\vec{S}_{n}+\Delta\sum_{n\sigma}{d_{n,xy,\sigma}^{\dagger}}{d_{n,xy,\sigma}}
\end{equation}
where $\lambda$ is the strength of the SOC and $\Delta$ describes the splitting of the $d_{xy}$ from $d_{yz,zx}$ due to a non-octahedral crystal field. $n$ is the index of the Ir sites. This Hamiltonian splits the  energy levels within the $t_{2g}$ manifold into three doublets,
\begin{equation}
E_{0,2}=\frac{\lambda}{4}(-1+\delta\mp\sqrt{9+2\delta+\delta^2}), E_1=\frac{\lambda}{2}
\end{equation}
where $\delta=\frac{2\Delta}{\lambda}$ and $E_0$ is the ground state energy. Our RIXS spectra determine the energy differences to be $E_1-E_0=0.58$ eV and $E_2-E_0=0.81$ eV, which leads to $\lambda=0.44$ eV and $\Delta=0.31$ eV. That is, the non-cubic crystal field is of a similar magnitude to SOC. With the parameters for $H_{local}$ in hand, the wavefunctions of the three doublets can now be determined. We find,
\begin{eqnarray}
\begin{array}{llll}
        |\phi_0\rangle &=& \frac{1}{1.56}(0.65d_{xy\uparrow,\downarrow}+\ri d_{yz\downarrow,\uparrow}\pm d_{zx\downarrow,\uparrow}) \\
        |\phi_1\rangle &=& \ \ \frac{1}{\sqrt{2}}(\ri d_{yz\uparrow,\downarrow}\pm d_{zx\uparrow,\downarrow}) \\
        |\phi_2\rangle &=& \frac{1}{2.20}[2d_{xy\uparrow,\downarrow}-0.65(\ri d_{yz\downarrow,\uparrow}\pm d_{zx\downarrow,\uparrow})] \\
\end{array}
\label{jdis}
\end{eqnarray}
The above effective Hamiltonian description is simple but informative. The three doublet picture naturally explains the data with a SOC strength $\lambda$ that is consistent with earlier calculations \cite{Mott1}. Importantly, however, it suggests a significant non-cubic crystal field, one that is comparable with the SOC - despite the small distortions of the octahedra in this material. This non-cubic crystal field strongly modifies the relevant wavefunctions. For example, $|\phi_0\rangle$ shows significant deviation from the ideal case: instead of the expected equal admixture of the three $t_{2g}$ orbitals in the ground state, the contribution from the $d_{xy}$ orbitals is reduced by more than $30\%$ in Sr$_3$CuIrO$_6$. This significant modification is illustrated in Fig.\,4. Given that anisotropic magnetic interactions lie at the heart of much of the proposed novel phenomena for the iridates, such a modification of the wavefunctions would be crucial and warrants further study.
\begin{figure}[ht]
\includegraphics[width=0.48\textwidth]{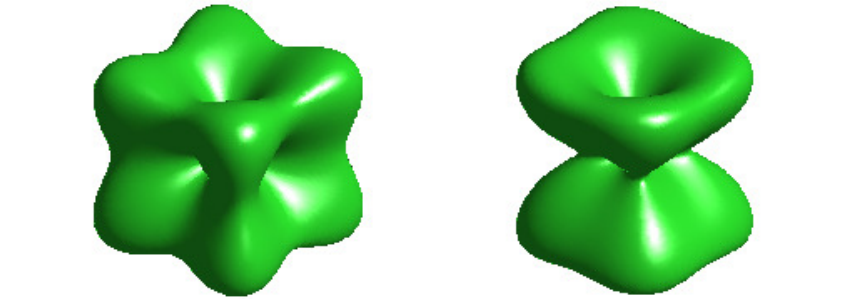}
\caption{(Color online) Density profile of the $t_{2g}$ hole. Left: $j_{\mathsf{eff}}=\frac{1}{2}$ state for perfect IrO$_6$ octahedra. Right: the modified wavefunction $|\phi_0\rangle$ (Eqn.\,3) due to the non-cubic crystal field arising from distortions of the octahedra.}
\label{fig4}
\end{figure}

To do so, we performed {\it ab initio} many-body calculations based on wavefunction electronic-structure theory. Complete-active-space self-consistent-field (CASSCF) and multireference configuration-interaction (MRCI) techniques from modern quantum chemistry \cite{QCmethod,QC2} were employed, as implemented in the {\sc molpro} package\cite{Mol}. Multiorbital and multiplet physics, SOC's, and O 2{\it p} to Ir 5{\it d} charge transfer effects are all treated on an equal footing, fully {\it ab initio}. The calculations were carried out on a cluster which contains a central IrO$_6$ octahedron, two nearest-neighbor CuO$_4$ plaquettes, and the adjacent Sr atoms, properly embedded in a large array of point charges that reproduces the crystal Madelung field in the cluster region.

The effect of non-cubic crystal fields is studied by first turning off the SOC. Without SOC, the embedded-cluster MRCI calculations give a splitting of 0.45 eV between the $d_{xz,yz}$ and $d_{xy}$ levels. This value is consistent with the $\Delta$ parameter derived in the model-Hamiltonian analysis carried out above. 

With spin-orbit interactions switched on \cite{SOC_molpro}, the MRCI calculations predict excitations within the $t_{2g}$ manifold at 0.59 and 0.91 eV.
At higher energies, we find Ir $t_{2g}^5$ to $t_{2g}^4e_g^1$ excitations between 3.2 and 4.9 eV and O $2p$ to Ir $5d$ charge transfer transitions between 3.5 and 5.4 eV. Further, excitation energies of $\sim$5.9 eV are predicted for the Ir $t_{2g}^3e_g^2$ states, see Table\,I.
\begin{table}
\caption{The percentage contributions of the different Ir {\it 5d$^5$} configurations to the different excitations. The results are obtained from MRCI+SOC calculations.}
\begin{ruledtabular}
\begin{tabular}{@{}c@{}c@{}@{}c@{}@{}c@{}@{}c@{}@{}c@{}@{}c@{}}
\backslashbox[2cm]{Config.}{Energy(eV)} & 0 & 0.59 & 0.91 & 3.22--4.84 & 5.88--5.91 \\
\hline
$d_{xy}^2d_{xz}^2d_{yz}^1$ & 37.1 & 49.1 & 13.8   &   &   \\
$d_{xy}^2d_{xz}^1d_{yz}^2$ & 49.0 & 47.8 & 2.4    &   &   \\
$d_{xy}^1d_{xz}^2d_{yz}^2$ & 13.9 &  3.1 & 83.8   &   &   \\

$t_{2g}^4 e_g^1$ &  &  &  & $>$96.0 &   \\
$t_{2g}^3 e_g^2$ &  &  &  &       & $>$99.0 \\
\end{tabular}
\end{ruledtabular}
\label{tab:tab1}
\end{table}

As seen in Fig.\,3, distinct spectral features are found by RIXS in all those energy windows. The good agreement between the calculated values and the experimental observations confirms the reliability of applying MRCI calculations in this context. In Table\,I, the contribution of the different orbital configurations to each state within the $t_{2g}^5$ manifold is listed. Again, those weights significantly deviate from the values corresponding to an undistorted octahedron, in agreement with the effective model. For example, both approaches predict that in the ground-state configuration the hole is preferentially distributed over the $d_{xz,yz}$ orbitals, due to a sizable splitting between the $d_{xz,yz}$ and $d_{xy}$ levels. The ground-state wavefunction $|\phi_0\rangle$ from the effective model displays a hole occupation ratio of $0.65^2$:1:1, i.e., 0.18:0.41:0.41. This compares very well to the percentages found in the MRCI+SOC calculations (see Table\,I, first column). The same holds for the other two doublets.

Our experimental and theoretical studies in Sr$_3$CuIrO$_6$ thus establish the importance of the local environment in determining the electronic structure of the Ir $t_{2g}$ manifold. With only a moderate distortion of the IrO$_6$ octahedra away from ideal cubic symmetry, the induced non-cubic crystal field has a similar strength to the SOC. This drives the system away from the strong SOC limit ($\frac{\lambda}{\Delta}>>1$) and significantly modifies the relevant wavefunctions, which will have important consequences for the magnetic exchange interactions. 

We now discuss the relevance of this work to other iridates, where to date, most of the theoretical work has assumed the strong SOC limit and treated the relevant electrons as in the effective $j=\frac{1}{2}$ states in which $d_{xy}$, $d_{xz}$ and $d_{yz}$ are equally mixed. Taking each of the iridate families in turn, we find for the pyrochlores $R_2$Ir$_2$O$_7$ that while all the O-Ir-O bonds are straight and of equal length, they rotate towards each other by 6$\sim$10$^\circ$ \cite{str227}.
In Na$_4$Ir$_3$O$_8$, the O-Ir and Ir-O bonds deviate from 90$^\circ$ by up to 9$^\circ$, and differ in length by up to 5\% \cite{hyperkagome2Y}. For Na$_2$IrO$_3$, the structure remains controversial. While a dramatic distortion was reported earlier\cite{honeycomb3}, recent results suggest it is smaller with the O-Ir-O bonds bending away from straight by $\sim$7$^\circ$ and the O-Ir and Ir-O bonds deviating from 90$^\circ$ by up to 10$^\circ$\cite{N213}. In all cases, the distortions are similar in magnitude or larger than the distortions present in Sr$_3$CuIrO$_6$. Thus treating these systems as being in the strong SOC limit is likely to be incorrect. With the electronic structure and magnetic interactions highly dependant on the relative contribution from the different orbitals, the non-octahedral crystal fields must be explicitly considered in any realistic models. Finally, we look at Sr$_2$IrO$_4$. In this system the octahedral bond angles are 90$^\circ$, and the bond lengths differ by only 4\% \cite{str214}. Thus this system may be the closest to the strong SOC limit, consistent with experimental observations \cite{Kim1}. However, the present study suggests that this may be the exception rather than the rule. 

In summary, we have used resonant inelastic x-ray scattering, a microscopic model Hamiltonian and $ab~initio$ quantum chemistry calculations to study the electronic excitations in Sr$_3$CuIrO$_6$, where Cu and IrO$_6$ octahedra form a 1-D spin chain. A three-level structure of the Ir $t_{2g}$ manifold is observed and ascribed to the lowering of the local octahedral symmetry. We show that this lowering of symmetry leads to a strong non-octahedral crystal field, which is comparable to the spin-orbit-coupling and modifies the electronic structure, the ground state wavefunctions and the magnetic exchange interactions significantly. It should therefore be taken into account when modeling the magnetic behavior of all non-ideal iridates. Important examples in the current literature include the pyrochlores $R_2$Ir$_2$O$_7$, the honeycomb lattice $A_2$IrO$_3$ and the hyperkagome $A_4$Ir$_3$O$_8$.

The work at Brookhaven was supported by the U.S. Department of Energy, Division of Materials Science, under Contract No. DE-AC02-98CH10886. The work at IFW Dresden was supported by the Computational Materials and Chemical Sciences Network program of the Division of Materials Science and Engineering, U.S. Department of Energy, through Grant No. DE-SC0007091. Use of the Advanced Photon Source was supported by the U. S. Department of Energy, Office of Science, Office of Basic Energy Sciences, under Contract No. DE-AC02-06CH11357. T.F.Q and G.C. were supported by the NSF through Grant No. DMR-0856234.


\begin{thebibliography}{1}
\bibitem{Mott1} B. J. Kim {\it et al.}, Phys. Rev. Lett., {\bf 101}, 076402 (2008).
\bibitem{Mott2} D. Pesin and L. Balents, Nature Physics, {\bf 6}, 376 (2010).
\bibitem{Mott3} G. Jackeli and G. Khaliullin, Phys. Rev. Lett., {\bf 102}, 017205 (2009).
\bibitem{honeycomb1} A. Shitade, H. Katsura, J. Kune\v{s}, X.-L. Qi, S.-C. Zhang and N. Nagaosa, Phys. Rev. Lett., {\bf 102}, 256403 (2009).
\bibitem{honeycomb2} J. Chaloupka, G. Jackeli and G. Khaliullin, Phys. Rev. Lett., {\bf 105}, 027204 (2010).
\bibitem{honeycomb3} Yogesh Singh and P. Gegenwart, Phys. Rev. B. {\bf 82}, 064412 (2010).
\bibitem{pyrochlore2} S. Nakatsuji {\it et al.}, Phys. Rev. Lett. {\bf 96}, 087204 (2006).
\bibitem{pyrochlore3} Yo Machida, Satoru Nakatsuji, Shigeki Onoda, Takashi Tayama and Toshiro Sakakibara, Nature {\bf 463}, 210 (2010).
\bibitem{hyperkagome2Y} Y. Okamoto, M. Nohara, Hiroko Aruga-Katori and H. Takagi, Phys. Rev. Lett., {\bf 99}, 137207 (2007).
\bibitem{hyperkagome2L} L. Balents, Nature, {\bf 464}, 199 (2010).
\bibitem{hyperkagome2M} M. J. Lawler, A. Paramekanti, Y. B. Kim, and L. Balents, Phys. Rev. Lett., {\bf 101}, 197202 (2008).
\bibitem{hyperkagome2Yi} Y. Zhou, P. A. Lee, T.-K. Ng, and F.-C. Zhang, Phys. Rev. Lett., {\bf 101}, 197201 (2008).
\bibitem{SC1} Jungho Kim {\it et al.}, Phys. Rev. Lett., {\bf 108}, 177003 (2012).
\bibitem{SC214} Fa Wang and T. Senthil, Phys. Rev. Lett., {\bf 106}, 136402 (2011).
\bibitem{Gang1} S. J. Moon {\it et al.}, Phys. Rev. Lett., {\bf 101}, 226402 (2008).
\bibitem{Gang2} M. A. Laguna-Marco {\it et al.}, Phys. Rev. Lett., {\bf 105}, 216407 (2010).
\bibitem{sizeU1} R. S. Singh {\it et al.}, Phys. Rev. B {\bf 77}, 201102(R) (2008).
\bibitem{sizeU2} Kalobaran Maiti, Solid State Commun. {\bf 149}, 1351 (2009).
\bibitem{mag327} J.W. Kim {\it et al.}, Phys. Rev. Lett., {\bf 109}, 037204 (2012).
\bibitem{M213} X. Liu {\it et al.}, Phys. Rev. B {\bf 83}, 220403(R) (2011).
\bibitem{N213} S.K. Choi {\it et al.}, arXiv:1202.1268.
\bibitem{TCF1} Choong H. Kim, Heung Sik Kim, Hogyun Jeong, Hosub Jin, and Jaejun Yu, arXiv:1201.5929.
\bibitem{TCF2} Subhro Bhattacharjee, Sung-Sik Lee and Yong Baek Kim, arXiv:1108.1806.
\bibitem{str3116} M. Neubacher and Hk.M\"{u}ller-Buschbaum, Z. Anorg. Allg. Chem. {\bf 607}, 124 (1992).
\bibitem{hybri} In some other systems the IrO$_6$ octahedra show larger hybridization with their neighbors than in Sr$_3$CuIrO$_6$. It is thus important in such cases to also clarify the significance of such effects (though they are negligible in the present case).
\bibitem{mag3116} Asad Niazi, P. L. Paulose, and E.V. Sampathkumaran, Phys. Rev. Lett., {\bf 88}, 107202 (2002).
\bibitem{RIXS} L. J. P. Ament, M. van Veenendaal, T. P. Devereaux, J. P. Hill, and J. van den Brink, Rev. Mod. Phys., {\bf 83}, 705 (2011); L. J. P. Ament, G. Khaliullin, and J. van den Brink, Phys. Rev. B {\bf 84}, 020403(R) (2011).
\bibitem{Kim1} B. J. Kim, T. Komesu, S. Sakai, T. Morita, H. Takagi and T. Arima, Science {\bf 323}, 1329 (2009).
\bibitem{QCmethod} T. Helgaker, P. J{\o}rgensen and J. Olsen, {\it Molecular Electronic-Structure Theory} (Wiley, Chichester, 2000).
\bibitem{QC2} L. Hozoi, L. Siurakshina, P. Fulde, and J. van den Brink, Sci. Rep. {\bf 1}, 65 (2011).
\bibitem{Mol} {\sc Molpro} 2009 quantum chemistry package, University of Cardiff, United Kingdom (http://www.molpro.net).
\bibitem{SOC_molpro} A. Berning, M. Schweizer, H.-J. Werner, P. J. Knowles and P. Palmieri, Mol. Phys. {\bf 98}, 1823 (2000).
\bibitem{str227} N. Taira, M. Wakeshima, and Y. Hinatsu, J. Phys.: Condens. Matter {\bf 13}, 5527 (2001).
\bibitem{str214} M. K. Crawford, M. A. Subramanian, R. L. Harlow, J. A. Fernandez-Baca, Z. R. Wang and D. C. Johnston, Phys. Rev. B {\bf 49}, 9198 (1994).
\end{thebibliography}
\end{document}